\documentclass{article}

\usepackage{microtype}
\usepackage{graphicx}
\usepackage{subfigure}
\usepackage{booktabs} 

\usepackage{hyperref}

\usepackage[accepted]{icml2019}

\usepackage{amsmath,amsfonts,bm}

\def\eqref#1{equation~\ref{#1}}

\def\1{\bm{1}}

\DeclareMathAlphabet{\mathsfit}{\encodingdefault}{\sfdefault}{m}{sl}
\SetMathAlphabet{\mathsfit}{bold}{\encodingdefault}{\sfdefault}{bx}{n}

\DeclareMathOperator*{\argmin}{arg\,min}

\icmltitlerunning{Submission and Formatting Instructions for ICML 2019}

\usepackage{url}
\usepackage{caption}
\usepackage{cleveref}
\usepackage{dsfont}
\usepackage{lipsum}
\usepackage{mwe}
\usepackage[final]{pdfpages}

\newcommand{\Lagr}{\mathcal{L}}

\newcommand{\ie}{\textit{i}.\textit{e}., }
\newcommand{\eg}{\textit{e}.\textit{g}. }
\newcommand{\FM}{$\text{F}_{1}\text{-score }$}
\newcommand{\err}{\epsilon}

\usepackage{booktabs}
\usepackage{authblk}

\icmltitlerunning{Prediction of Workplace Injuries}

\begin{document}

\twocolumn[
\icmltitle{Prediction of Workplace Injuries}

\icmlsetsymbol{equal}{*}

\begin{icmlauthorlist}
\icmlauthor{Mehdi Sadeqi}{cority}
\icmlauthor{Azin Asgarian}{gpart}
\icmlauthor{Ariel Sibilia}{cority}
\end{icmlauthorlist}

\icmlaffiliation{cority}{Cority Inc, Toronto, Ontario, Canada}
\icmlaffiliation{gpart}{Georgian Partners Inc, Toronto, Ontario, Canada}

\icmlcorrespondingauthor{Mehdi Sadeqi}{Mehdi.Sadeqi@cority.com}

\icmlkeywords{Machine Learning, Causal Inference, Cost-Sensitive Learning, Transfer Learning, Imbalanced Data, Cost-Curves, Actionable Insights, Partial Dependence Plot}

\vskip 0.4in]
\printAffiliationsAndNotice{}

\begin{abstract}
Workplace injuries result in substantial human and financial losses. As reported by the International Labour Organization (ILO), there are more than 374 million work-related injuries reported every year. In this study, we investigate the problem of injury risk prediction and prevention in a work environment. While injuries represent a significant number across all organizations, they are rare events within a single organization. Hence, collecting a sufficiently large dataset from a single organization is extremely difficult. In addition, the collected datasets are often highly imbalanced which increases the problem difficulty. Finally, risk predictions need to provide additional context for injuries to be prevented. We propose and evaluate the following for a complete solution: 1) several ensemble-based resampling methods to address the class imbalance issues, 2) a novel transfer learning approach to transfer the knowledge across organizations, and 3) various techniques to uncover the association and causal effect of different variables on injury risk, while controlling for relevant confounding factors.
\end{abstract}

\vspace{-.4cm}
\section{Introduction}
Workplace injuries can affect workers' lives and can cause substantial economic burden to employees, employers, and more generally to society \cite{ILO,sarkaretal}. There are more than $374$ million work-related injuries reported every year, resulting in more than $2.3$ million deaths annually \cite{ILO}. The yearly cost to the global economy from work-related injuries alone is a staggering \$3 trillion, estimated by ILO. 

Predicting injuries and providing actionable insights on factors associated with injuries are critical for improving workplace safety. Recent research has focused on this problem in sports \cite{naglah2018athlete, rossi2018effective}, construction \cite{tixier2016application, poh2018safety}, and various workplace settings \cite{sanchez2011prediction, rivas2011explaining, sarkaretal}. Despite introducing many interesting frameworks, these studies do not address some of the main challenges such as lack of labeled data and class imbalance issues. In addition, previous works do not investigate the causal relationships between different variables and injury incidents.

We propose a framework that employs ensemble-based resampling methods and a novel transfer learning approach to address class imbalance and data availability issues. We apply a method to predictive features of injuries to highlight their direct causal effect and we utilize a visualization technique that provides interpretability. We demonstrate the utility of our framework through experiments performed on real-world datasets. More specifically, we show that ensemble-based resampling and transfer learning techniques can increase the $\text{F}_{1}\text{-score}$ by 100\% and area under precision recall curve by 44\%, when compared to a model trained on a single organization dataset. 

In the remainder of this paper, we first provide a brief overview of the problem and our machine learning framework in Section \ref{subsec:DataandProblemDescription}. We present the employed ensemble-based resampling techniques, our instance-based transfer learning method, and the approaches used to provide actionable insights in Sections \ref{subsec:ImbalancedData}, \ref{subsec:TransferLearning}, and \ref{subsec:ActionableInsights}, respectively. Section \ref{sec:Experiments} describes our results and Section \ref{sec:Conclusion} covers conclusions and future work.

\vspace{-.2cm}
\section{Injury Prediction as Supervised Learning}
\subsection{Data and Problem Description}
\label{subsec:DataandProblemDescription}
To conduct this study, we collected employees' safety-related information from different organizations during years 2016-2017. We treat the learning problem as a binary classification task. Using the data collected during 2016, the objective is to predict whether an employee was injured or not in 2017. The collected datasets differ in size and distribution, however, they are all highly imbalanced (1-7\% injury cases). In all datasets, the employee records are represented by 38 engineered features that capture two main groups of information: general employee information (\eg age), and event-based information. Event-based information are either associated with the employee (\eg number of absences) or with the employee's site (\eg the risk assessments scores).\footnote{The names of the organizations are masked due to confidentiality reasons.} In this work, we use XGBoost \cite{friedman2001greedy} as our base predictive model.

\subsection{Imbalanced Data}\label{subsec:ImbalancedData}
To address the problem of highly imbalanced data, several approaches are proposed in the literature. Among the most common ones are over-sampling and under-sampling methods \cite{chawla2003c4}, neighbor-based techniques \cite{Wilson:1972, Tomek:1976}, Synthetic Minority Over-sampling TEchnique (SMOTE) \cite{Chawla:2002}, adjusting class weights, boosting techniques, and anomaly detection methods. From these solutions, we are particularly interested in four methods that combine ensemble-based supervised learning algorithms with resampling methods (\textit{UnderBagging}, \textit{SMOTEBagging}, \textit{RUSBoost}, and \textit{SMOTEBoost}). We give a brief overview of these methods below.

UnderBagging and SMOTEBagging methods try to rebalance the class distribution in each bag of the bagging algorithms. UnderBagging uses random under-sampling while SMOTEBagging uses SMOTE or over-sampling to achieve this goal \cite{Galar:2012}. Alternatively, RUSBoost \cite{Seiffert:2010} and SMOTEBoost \cite{Chawla:2003} combine AdaBoost.M2 \cite{Freund:1997} boosting algorithms with resampling methods to address the class imbalance issues. Similar to UnderBagging and SMOTEBagging, in each iteration of training weak learners, these two algorithms respectively use random under-sampling (to reduce majority instances) and SMOTE (to increase minority instances). Moreover, these four approaches have the advantage of having very few number of hyper-parameters. We provide a comparison of these methods in Section \ref{subsec:ResImbalancedData}.

\subsection{Transfer Learning} 
\label{subsec:TransferLearning}
To handle the data unavailability issues for a new organization (target domain), we leverage the knowledge learned from other organizations (source domain) by employing an instance-based transfer learning method. Given a set of target training samples \scalebox{0.9}{$\left\{(x_i,y_i)| i \in \left\{1,2, ...,N_T\right\}\right\}$} and a loss function \scalebox{0.9}{$\Lagr(.)$} the goal of supervised learning is to find model \scalebox{0.9}{$\mathcal{A}^*$} that minimizes the expected error, \ie \scalebox{0.9}{$\mathcal{A}^* = \underset{\mathcal{A}\in\mathbb{A}}\argmin \mathbb{ E}_{x\sim P_T}\big[\Lagr(\mathcal{A}(x), y)\big]$.} Here $x$ is an arbitrary sample and \scalebox{0.9}{$P_T$} is the probability distribution of target samples. Following the idea of importance sampling \cite{liu2008monte, asgarian2018hybrid} for transferring the knowledge from source domain (\scalebox{0.9}{$S$}) to target domain (\scalebox{0.9}{$T$}), we can express the expected error as \scalebox{0.9}{$\alpha\mathbb{E}_{x\sim P_T}\Big[\err(x)\Big] + (1-\alpha)\mathbb{E}_{x\sim P_S}[\err(x)\tfrac{P_T(x)}{P_S(x)}]$}. Here \scalebox{0.9}{$\err(x)$} shows the error for each sample and $\alpha$ is a hyper-parameter that controls the overall relative importance between source and target samples. 
Source sample weights \scalebox{0.9}{$\{w_{x_j}=\frac{P_T(x_j)}{P_S(x_j)}\mid j\in\{1, ...,N_S\}\}$} play a major role in instance-based transfer learning methods, as they control the individual effect of source samples~\cite{asgarian2017subspace}. We describe different weighting approaches including our five baselines models and our proposed weighting strategy in the following.

\textbf{Baselines:} Models \scalebox{0.9}{$\mathcal{A}_S$}, \scalebox{0.9}{$\mathcal{A}_T$}, and \scalebox{0.9}{$\mathcal{A}_{S\cup T}$} trained respectively on source, target and the union of source and target, serve as minimum baselines that a transfer learning method must outperform. Our fourth baseline model is an instance-weighted model (\scalebox{0.9}{$\mathcal{A}_\mathbf{1}$}) with all the weights set to $\mathbf{1}$ (\ie \scalebox{0.9}{$\mathbf{W}_S = \mathds{1}$}). This is similar to \scalebox{0.9}{$\mathcal{A}_{S\cup T}$}, except in this model we use $\alpha$ to determine the relative overall importance between source and target samples. Our last baseline model \scalebox{0.9}{($\mathcal{A}_G$)}, assumes Gaussian distributions for target and source samples to evaluate the source sample weights $w_{x_j}$.

\textbf{Hybrid Weights:} Previous methods evaluate the source sample weights solely based on their similarity to the target domain. We argue that it is also important to measure the relevance of source samples to the target task. Hence, we define weights $w_{x}=w_{domain_x} + w_{task_x}$, where $w_{domain_x}$ measures the similarity of an arbitrary source sample $x$ to the target domain, while $w_{task_x}$ measures the importance of sample $x$ in the target task. 

For evaluating $w_{domain_x}$, unlike the previous methods that employ generative approaches to estimate \scalebox{0.9}{$P_T$} and \scalebox{0.9}{$P_S$}, we directly approximate weights \scalebox{0.9}{$w_{domain_x} = \frac{P_T(x)}{P_S(x)}$} with a discriminative classifier. More specifically, using \scalebox{0.9}{$\{(x, l_x)\ | x \in S\cup T \text{, and $l_x=1$ if $x \in S$, $l_x=0$ otherwise}\}$}, we train a binary classifier (e.g., logistic regression (LR)) to differentiate source and target samples. Next, we use the learned weights of this classifier \scalebox{0.9}{($w_{lr}$ and $c_{lr}$)} to estimate source sample weights \scalebox{0.9}{$w_{domain_x}=\frac{P_T(x)}{P_S(x)} \approx \frac{1}{\exp(x^T w_{lr} + c_{lr})}$}. 

To compute $w_{task_x}$, we train an instance of our predictive model \scalebox{0.9}{(XGBoost)} using all samples from source and target with their corresponding injury labels. We then define $w_{task_x}$ to be the uncertainty of this model about sample $x$. We define the uncertainty of a model about sample $x$ to be the distance of $x$ to the decision boundary. Note that this value could be negative (thus subtracting from $w_{x}$) when the decision is incorrect, or positive (thus adding to $w_{x}$) when the decision is correct. We denote this model as \scalebox{0.9}{$\mathcal{A}_{HW}$}. 

\subsection{Actionable Insights}
\label{subsec:ActionableInsights}

To visualize the relationship between injuries and its predictors, we explore one method to show the straightforward association, and one to find causal relationship.\footnote{Please note that in order to infer meaningful and accurate causal relationships using this approach, we need adequately accurate models. The ensemble-based resampling methods and transfer learning are an attempt in this direction.}

\begin{figure}[ht!]
\centering
\begin{minipage}{0.45\textwidth}
    \centering
    \includegraphics[width=1\textwidth]{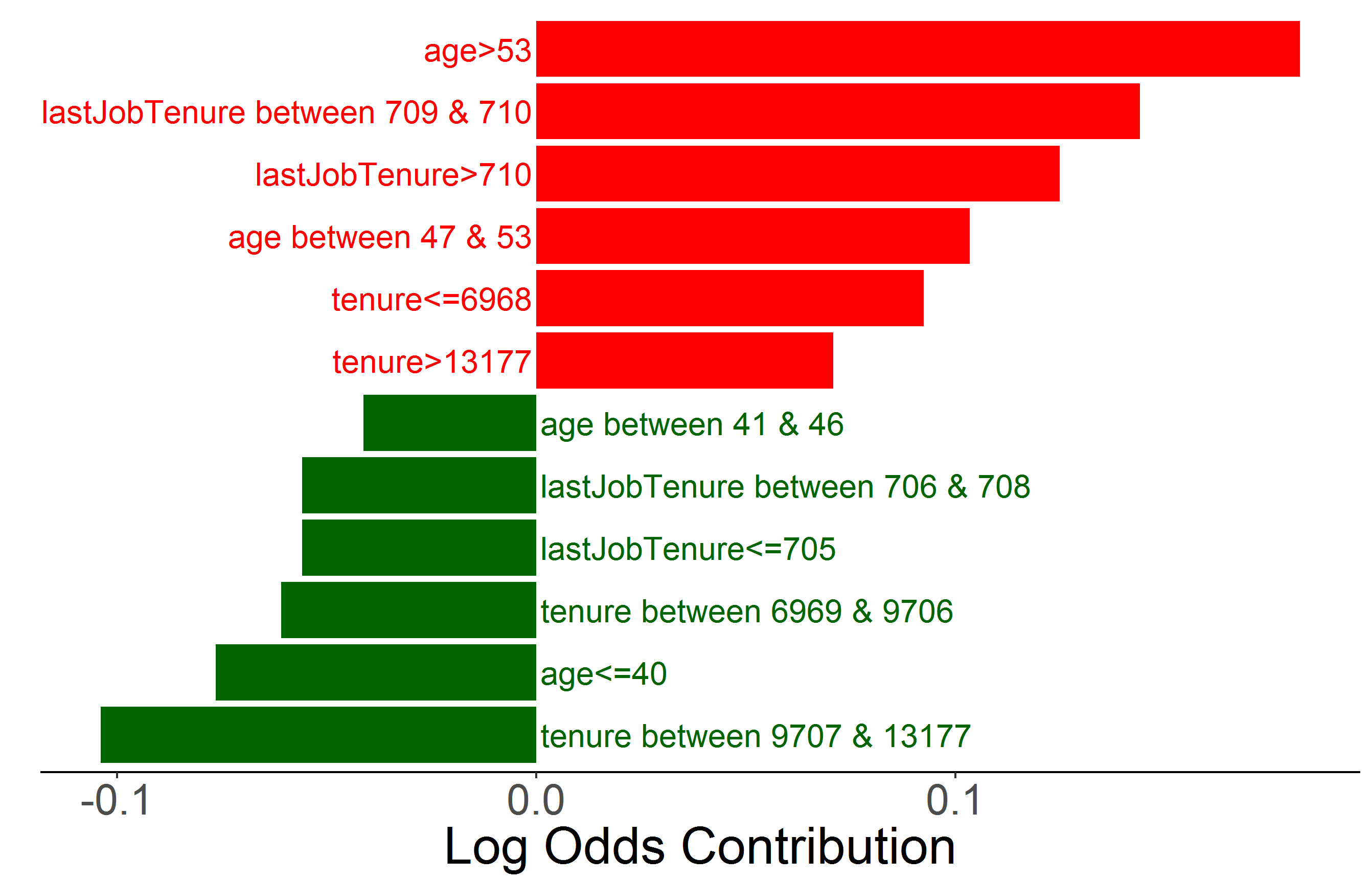} 
    \caption{\small Log-odds contribution of binned continuous variables for a hypothetical example company.}
    \label{fig:impact}
\end{minipage}
\vspace{-.2cm}
\end{figure}

To measure the association between features and the target variable, we average the contribution of each feature's possible values to the log-odds ratio of all samples which match that value. We bin every continuous variable, treating them as categorical, so that such matching is possible for all variables. We use the \textit{xgboost\_explainer} package in Python, which is inspired by \cite{xgbExplainer}, to find the average log-odds contribution of each feature to each sample. Next, for each discrete value of each feature, we average the sample-based contribution over all samples with matching values. This gives us a visualization of both the average impact and direction of each variable as seen in Figure \ref{fig:impact}. This is a unique approach for visualizing the association between safety-related variables and injuries and it also gives an interpretation of the model. However, it must be noted that each effect size measured here is an average over the population, not a deterministic effect. 

Partial dependence plots (PDPs) \cite{Friedman:2001} are an especially interesting and fairly natural way of adding interpretability to more complex models. PDPs show the average relationship between two variables over a population by marginalizing over the distribution of all other variables. For a trained model, this is approximated by summing over the training data, where, unlike the marginalized variables, the variables to be plotted are held constant \cite{Friedman:2001}. PDPs may also yield a causal interpretation of the effect of a variable on injuries \cite{Zhao:2017}. \citeauthor{Zhao:2017} showed that a partial dependence calculation that averages over a set of variables is equivalent to controlling for those variables using Pearl's back-door adjustment formula \cite{pearl1993}. For example, there are instances where the relationship between an input variable and injuries is reversed when doing a partial dependence calculation over another variable. In \cite{pearl2014}, Pearl shows that instances of the Simpson's paradox can be properly explained when using the back-door criterion to adjust for a variable.

\begin{figure}[ht!]
\centering
\begin{minipage}{0.43\textwidth}
    \centering
    \includegraphics[width=1\textwidth]{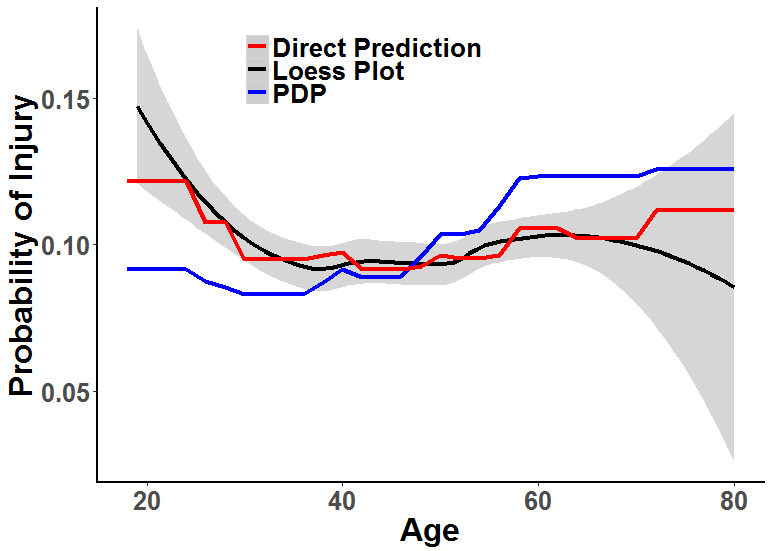} 
    \caption{\small  The loss curve and model prediction show the direct association between age and injury. The PDP curve shows the same relationship when controlling for tenure.}
    \label{fig:visual}
\end{minipage}
\vspace{-.2cm}
\end{figure}

In our example, we observe the effect of age on probability of injuries after adjusting for tenure. In our causal hypothesis, both age and tenure have a causal impact on injuries. However, neither are considered to directly cause each other. Nevertheless, they are strongly correlated, so we connect each of their exogenous variables in a causal graph. Intuitively, this can be described as both age and tenure being caused by the ``passage of time". In our causal path from age to injury there is one back-door path, which is blocked by tenure. Therefore, tenure satisfies the back-door criterion from age to injury. In Figure \ref{fig:visual}, we plot probability of injury versus age generated in three different ways. The direct prediction and loess (locally estimated scatterplot smoothing) plot both estimate the direct association between age and probability of injury. The partial dependence plot shows the same association, after adjusting for tenure.

\vspace{-.2cm}
\section{Experiments}
\label{sec:Experiments}
\subsection{Imbalanced Data}
\label{subsec:ResImbalancedData}
For comparing the ensemble-based resampling methods to our benchmark XGBoost model, we adopted an effective visualization technique called cost-curves \cite{Drummond:2006}. These curves allow us to evaluate a classifier in deployment conditions of two important factors--\textit{class distributions} and \textit{misclassification costs}--which are usually unknown or varying with time. Using cost-curves, we can visualize a classifier's performance for the whole range of these unknown factors. Receiver operating characteristic (ROC) Curves are point/line dual with cost-curves and they convey the same information implicitly, but they are not visually as informative. Hence, we used cost-curves in addition to our other evaluation metrics. These curves also helps to find the conditions for which a classifier shows better performance compared to other classifiers and particularly to trivial classifiers \cite{Drummond:2006}. 

\begin{figure}[ht!]
\vspace{-.1cm}
\centering
    \includegraphics[height=3.8cm, width=6.7cm]{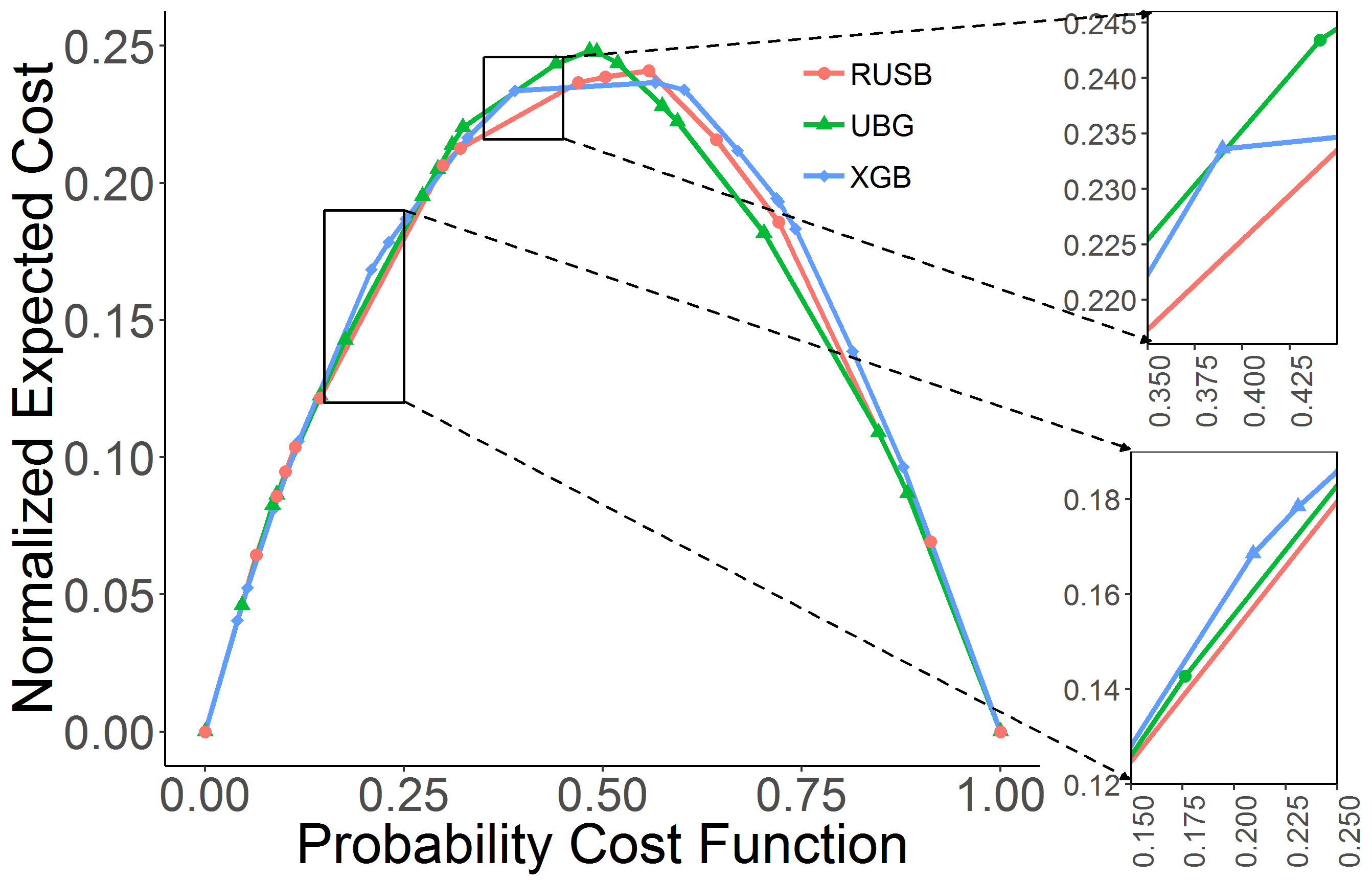}
\caption{\small Cost-curves of RUSBoost and UnderBagging versus XGBoost.}
\vspace{-.2cm}
\label{fig:Cost-Curves}
\end{figure}

In all datasets, RUSBoost and UnderBagging showed a better performance than the XGBoost model in handling class imbalance (Figure \ref{fig:Cost-Curves}). SMOTEBagging and SMOTEBoost, however, showed a lower performance compared to the XGBoost. To avoid a cluttered plot, they are not shown in Figure \ref{fig:Cost-Curves}. That being said, this is a data-dependent behavior and one should test each of these algorithms to see which one best matches the data. 

\begin{figure}[ht!]
\vspace{-.2cm}
\centering
    \includegraphics[height=3.3cm, width=6.2cm]{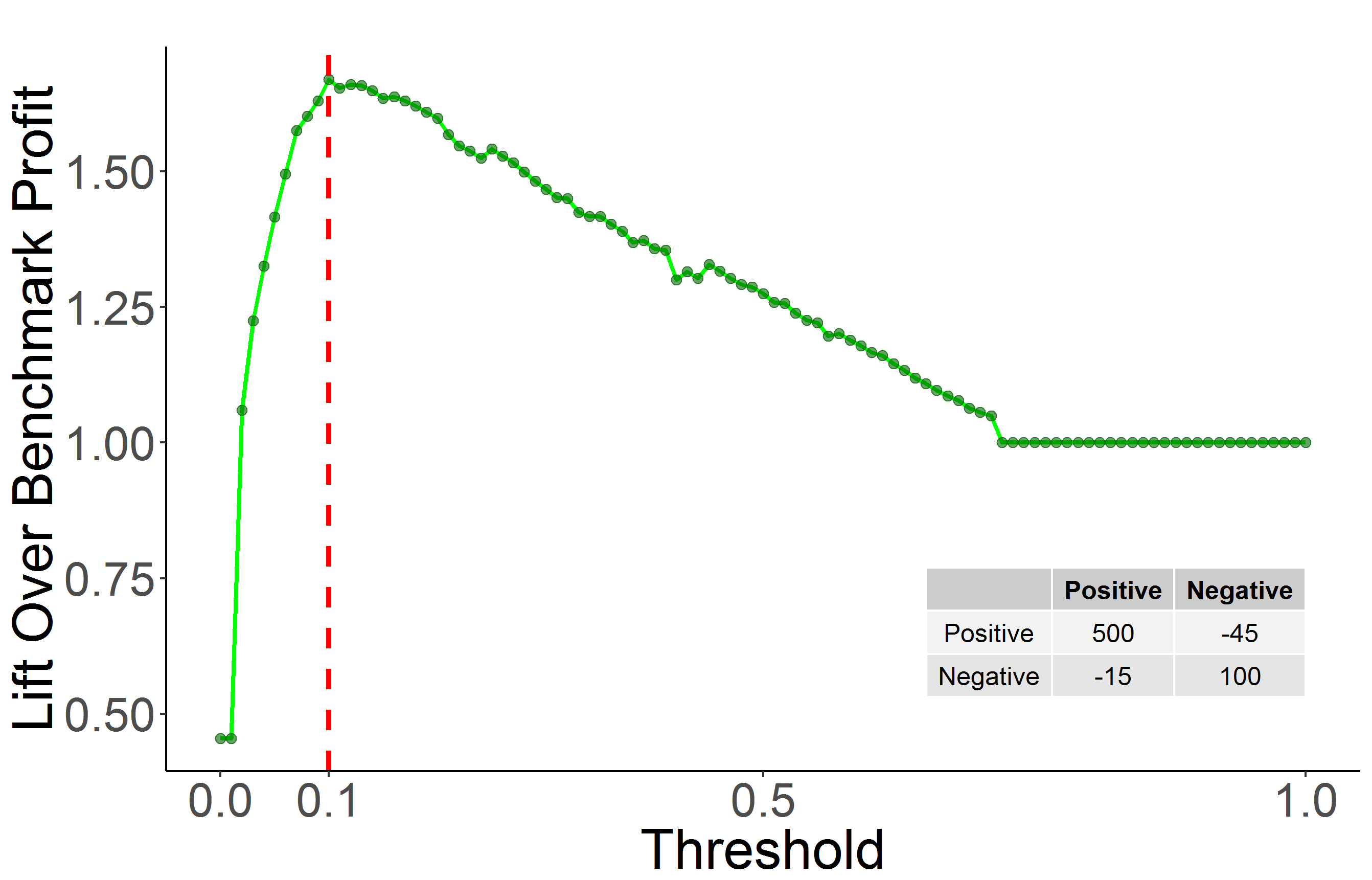}
\caption{\small Optimum threshold for a given profit matrix.}
\vspace{-.5cm}
\label{fig:Cost-Sensitive}
\end{figure}

The cost-curve performance comparison and model selection is only needed if misclassification costs are unknown in advance or deployment class distribution is different from the test data class distribution. If this is not the case, we can find the optimum threshold that maximizes a profit function defined by a profit matrix. This threshold will be another hyper-parameter that should be optimized inside a cross-validation pipeline. In our XGBoost model, we used the profit function as the evaluation metric for \textit{early stopping} for $100$ different thresholds. For simplicity, here we kept all other XGBoost parameters fixed and optimized only for threshold. Figure \ref{fig:Cost-Sensitive} shows the ratio of model profit to a benchmark profit as a function of threshold values. The profit matrix and the optimum threshold value $0.1$ are shown in this figure.

\subsection{Transfer Learning}
In our transfer learning framework, we considered Organization-1's dataset as the target domain and Organization-2's dataset as the source domain. For training, we used 58,271 samples from target (12,225) and source (46,046) training sets, and evaluated the models on 3,057 samples from target test set. Since the datasets were highly imbalanced (1-7$\%$ injury cases), we used precision, recall, \FM (macro), and area under the precision-recall curve (AUCPR) as our four evaluation metrics. Results of our quantitative evaluation is shown in Table \ref{table:TL}. 

In Table \ref{table:TL}, we see that model $\mathcal{A}_T$ has a poor performance with \FM  equal to 0.06 and AUCPR of 0.0375. This is possibly due to data sparsity issue and lack of expressiveness of the model. On the other hand, $\mathcal{A}_S$ has a higher \FM and AUCPR, but the precision is diminished. Also model $\mathcal{A}_{S\cup T}$ performs better in terms of \FM and AUCPR compared to both models $\mathcal{A}_T$ and $\mathcal{A}_S$. The best result is obtained with our model $\mathcal{A}_{HW}$, which increases the \FM and AUCPR considerably. 

\begin{table}[ht!]
\centering
\begin{tabular}{ccccc}
\toprule
Method & Precision & Recall & $\text{F}_{1}\text{-score}$ & AUCPR\\
\midrule
$\mathcal{A}_{T}$              & 0.07 & 0.06 & 0.06 & 0.0375\\ 
$\mathcal{A}_{S}$              & 0.04 & 0.18 & 0.07 & 0.0405\\ 
$\mathcal{A}_{S\cup T}$            & 0.13 & 0.06 & 0.08 & 0.0478\\ 
$\mathcal{A}_\mathbf{1}$       & 0.06 & 0.12 & 0.08 & 0.0456\\ 
$\mathcal{A}_{G}$              & 0.07 & 0.16 & 0.10 & 0.0532\\ 
$\mathcal{A}_{HW}$            & 0.11 & 0.12 & \textbf{0.12} & \textbf{0.0542}\\
\bottomrule
\end{tabular}
\caption{\small Performance of different methods on Company-1's data.}
\vspace{-.1cm}
\label{table:TL}
\end{table}

\begin{figure}[ht!]
\vspace{-.3cm}
\centering
\includegraphics[width=.35\textwidth]{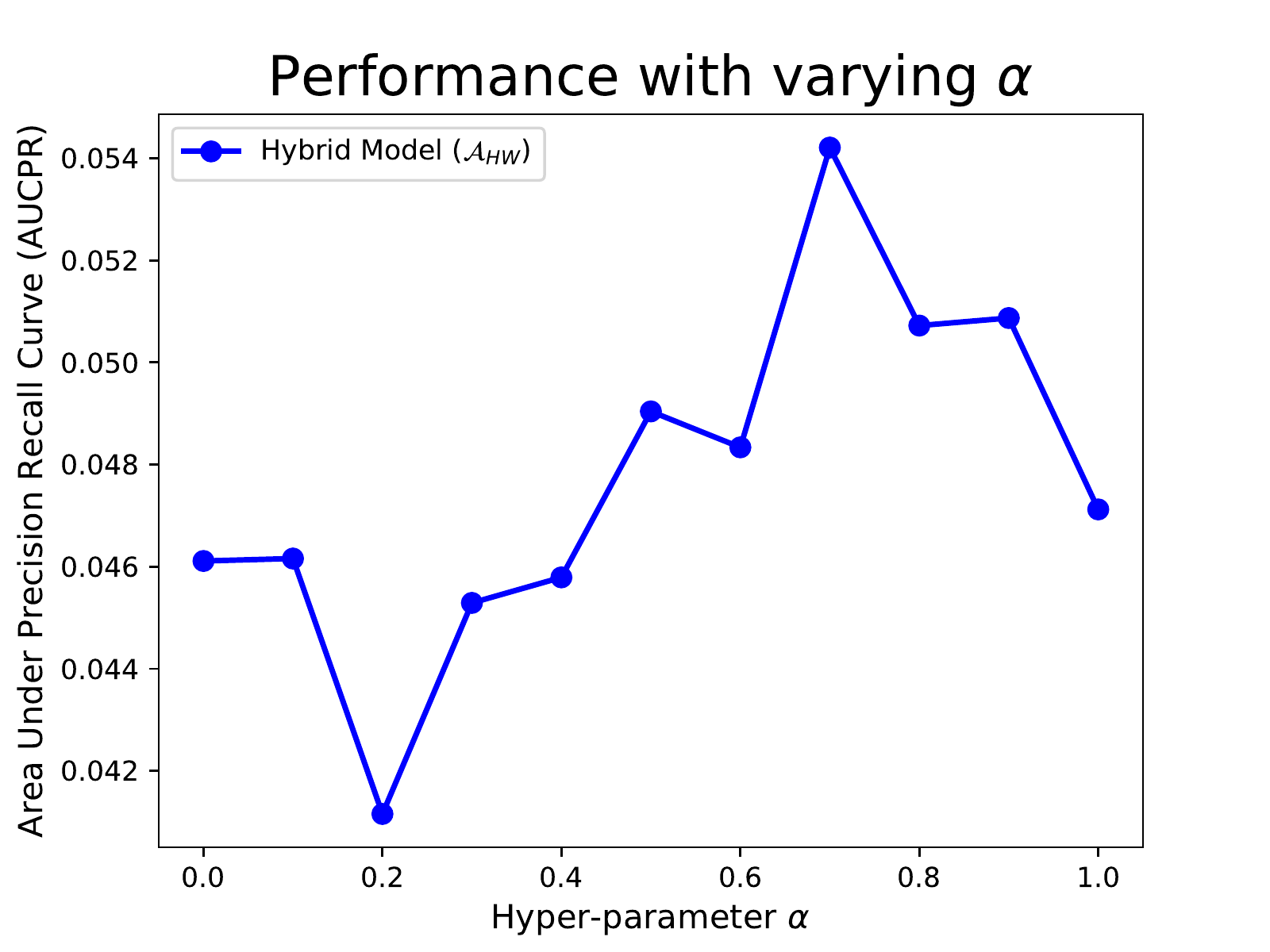}
\caption{\small Effect of $\alpha$ on performance.}
\label{fig:TLalpha}
\vspace{-.5cm}
\end{figure}

Figure \ref{fig:TLalpha} shows the AUCPR obtained with model $\mathcal{A}_{HW}$ as a function of hyper-parameter $\alpha$. We see that the best performance is achieved with $\alpha$ equal to 0.7. However, increasing or decreasing $\alpha$ results in lower AUCPR, as it enhances the influence of target or source samples respectively.

\section{Conclusions And Future Work} 
\label{sec:Conclusion}
In this paper, we investigate the problem of injury risk prediction in a supervised learning framework. To improve the performance in presence of highly imbalanced data, we employ ensemble-based resampling techniques. To address the lack of labeled data, we propose an instance-based transfer learning method. Additionally, we provide actionable insights to prevent injuries and show the effectiveness of our framework experimentally. In our future work, we focus further on discovering the causal relationships from observational data as it is a key element in injury prevention. 

\section*{Acknowledgment} 
\label{sec:Acknowledgment}
We would like to acknowledge Parinaz Sobhani, Madalin Mihailescu, Chang Liu, and Diego Huang for their invaluable assistance and insightful comments on the initial draft of this work. We are specially grateful to our management team at Georgian Partners and Cority including Madalin Mihailescu, Ji Chao Zhang, Stan Marsden, and David Vuong for encouraging and supporting our research activities, without which we were unable to complete this work. Finally, we would like to thank Babak Taati for his direction and guidance on this study. 

\bibliography{egbib}
\bibliographystyle{icml2019}

\includepdf[pages=1]{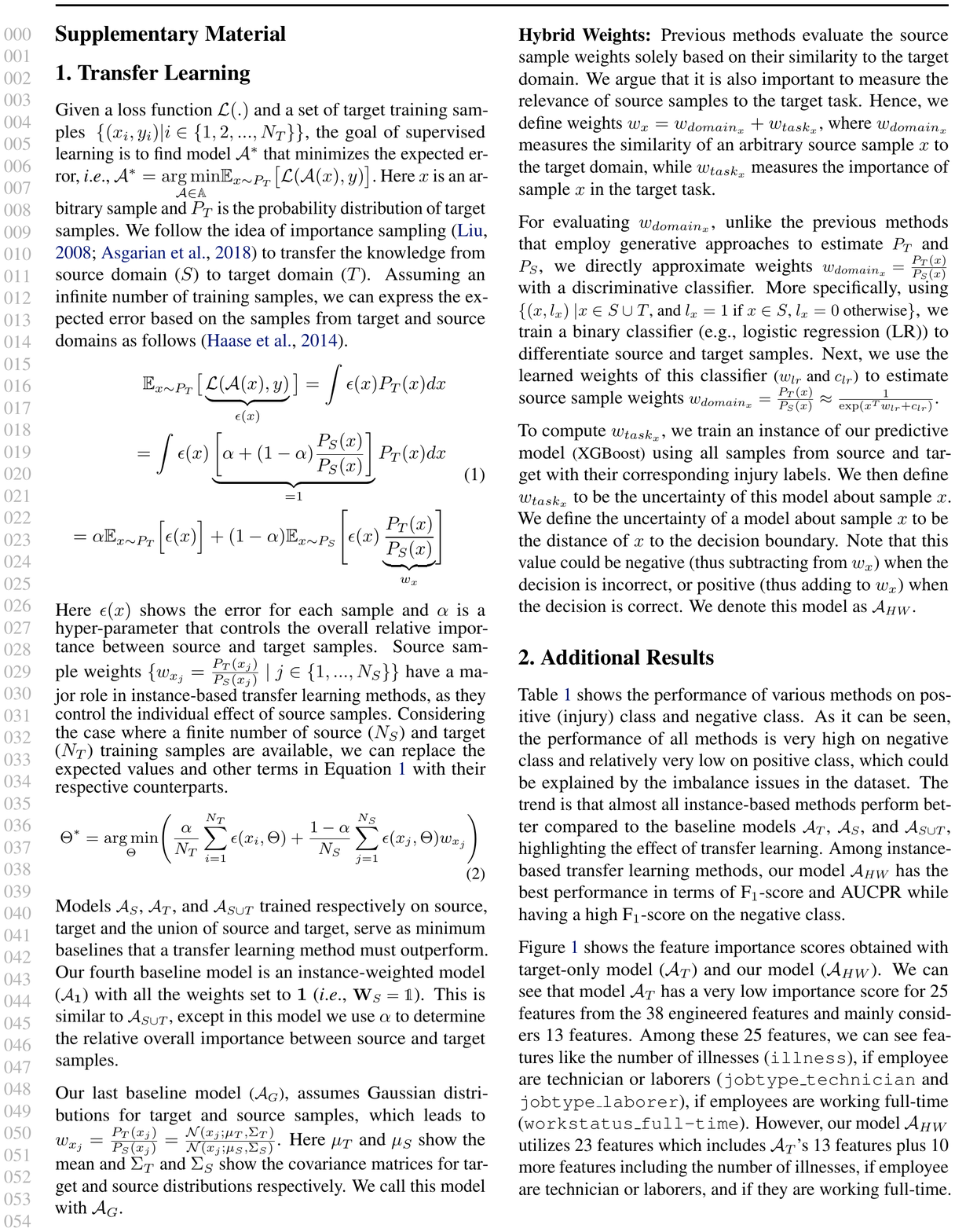}
\includepdf[pages=2]{supp.pdf}
\end{document}